# Quantum Algorithm for Jaccard Similarity

Varun Puram
Dept of Computer Science
Oklahoma State University
Stillwater,USA
vpuram@okstate.edu

Ruthvik Rao Bobbili
Dept of Computer Science
Oklahoma State University
Stillwater,USA
rbobbil@okstate.edu

Johnson P Thomas
Dept of Computer Science
Oklahoma State University
Stillwater,USA
jpt@cs.okstate.edu

***Abstract***: -Jaccard Similarity is a very common proximity measurement used to compute the similarity between two asymmetric binary vectors. Jaccard Similarity is the ratio between the $\#_{1's}$ (Intersection of two vectors) to $\#_{1's}$ (Union of two vectors). This paper introduces a quantum algorithm for finding the Jaccard Similarity $\#_{1's}$, in the Intersection and Union of two binary vectors. There are two sub-algorithms one for each. Measuring the register for respective algorithm gives count of number of 1's in binary format. Implementation on IBM composer is also included.

## Introduction

Quantum computing is based on quantum mechanics to perform computations. In recent years, quantum computing has emerged as an important research area where quantum computing is applied to many classical tasks to reduce the time and space complexity. The biggest breakthrough of Quantum computing is Shor's Algorithm [1] which was proposed in 1994 by the American Mathematician Peter Shor. These algorithms is exponentially faster than the most efficient classical algorithm. Grover's search [2] algorithm is used to search an element in an unsorted array with time $O(\sqrt{N})$ time which is much faster than all the classical search algorithms. Solving quantum algorithms in quantum computer is much faster than classical super computers. Even Quantum Machine Learning tasks are now showing significant computational speed up in their performances when compared with classical computing [3]-[4] tasks.

## Related Work

Some quantum algorithms to measure the similarity between two binary vectors have been proposed in recent years. The Quantum Binary Classifier paper [5] proposed similarity between vectors using Cosine Similarity. The set of input vectors are trained using labels, and the cosine similarity between the testing and trained vectors are determined using the dot product of the test and trained vectors. If the cosine similarity is high then the respective label appended to the test vector. Quantum K nearest neighbor based on the hamming distance is proposed in [6], In this paper the hamming distance between two binary vectors is used to find the nearest neighbor. The Hamming distance is essentially the number of bit flips from one vector to another. The Jaccard Similarity is one of the proximity measurements for finding the similarity between two asymmetric binary vectors. It considers 1-1 matches and ignores 0-0 matches in the vectors. As far as we are aware, no quantum algorithm has been proposed for the Jaccard similarity. In this paper we propose a quantum model for the Jaccard similarity between two binary vectors.

## Methodology

Initially we solve the Jaccard Similarity using the classical method.

Let $\vec{x}$ and $\vec{y}$ are two binary vectors. The Jaccard Similarity is calculated using classical computations as:

$$J(\vec{x},\vec{y}) = \frac{\#_{1's}(\vec{x} \cap \vec{y})}{\#_{1's}(\vec{x} U \vec{y})} \quad (1)$$

For example, $\vec{x}$ = 100011011100, $\vec{y}$ = 110011100100 then $\vec{x} \cap \vec{y}$ = 100011000100 and $\vec{x}U\vec{y}$ = 110011111100

$$J(\vec{x},\vec{y}) = \frac{\#_{1's}(100011000100)}{\#_{1's}(110011111100)} \quad (2)$$

$$J(\vec{x},\vec{y}) = \frac{4}{8} \quad (3)$$

Like the cosine measure, the Jaccard Similarity ignores 0-0 matches. For the similarity check for two asymmetric binary vectors; Jaccard Similarity is one of the popular ways to calculate the similarity. Here we describe two sub-Algorithms for calculating $\#_{1's}(\vec{x} \cap \vec{y})$ and $\#_{1's}(\vec{x} \cup \vec{y})$.

**1) Quantam Method for $\#_{1's}(\vec{x} \cap \vec{y})$ :**

Let $\vec{x}$ and $\vec{y}$ are two binary vectors, $\vec{x},\vec{y} \in \mathbb{R}^N$, where we need to find $\#_{1's}$ which is the intersection between them. Prepare the state with three registers.

$$|\varphi\rangle = |x\rangle \otimes |y\rangle \otimes |a\rangle \quad (4)$$

Where $|x\rangle$ = $|x_{N-1} \, x_{N-2} \, x_{N-3}\ldots x_0\rangle$ = $|x_{N-1}\rangle| \, x_{N-2}\rangle|x_{N-3}\rangle\ldots|x_0\rangle$ = $|x_{N-1}\rangle \otimes |\, x_{N-2}\rangle \otimes |x_{N-3}\rangle\ldots \otimes|x_0\rangle$;

$|y\rangle$ = $|y_{N-1} \, y_{N-2} \, y_{N-3}\ldots \, y_0\rangle$ = $|y_{N-1}\rangle| \, y_{N-2}\rangle|y_{N-3}\rangle\ldots|y_0\rangle$ = $|y_{N-1}\rangle \otimes |\, y_{N-2}\rangle \otimes |y_{N-3}\rangle\ldots \otimes|y_0\rangle$.

A binary vector $\vec{a} \in \mathbb{R}^m$ can be used to store integer value of the count of $\#_{1's}$ in the intersection in binary format, the count means numbers of 1's present in the intersection of $\vec{x},\vec{y}$.

$\vec{a}$ = $|a_m \, a_{m-1}\ldots a_0\rangle$ and we get the integer value by

$$a = a_m 2^m + a_{m-1} 2^{m-1}\ldots + a_0 2^0 ; \quad (m = \lfloor logN + 1 \rfloor) \quad (5)$$

Apply Algorithm 1 intersection $(\vec{x},\vec{y},\vec{a})$ shown below and measure the register $|a\rangle^{\otimes m.}$

Intersection $(\vec{x}, \vec{y}, \vec{a})$:

1) **Input:** Vectors $\vec{x} = (x_{n-1}x_{n-2}...x_0)$, $\vec{y} = (y_{n-1}y_{n-2}....y_0)$
2) **Output:** $a = \#_{1's}(\vec{x} \cap \vec{y})$
3) Initialize the registers $Q_x^{\otimes N} Q_y^{\otimes N} Q_a^{\otimes m}$
4) *for* all $x_i$ and $y_i$:
   - if ($x_i = 1$):
     - apply X gate to $x_i$ qubit
   - end if
   - if ($y_i = 1$):
     - apply X gate to $y_i$ qubit
   - end-if
5) end for
6) for i=0 to N-1:
   - for j=0 to m:
     - if (m-j-1) ≥ 0:
       - Apply multi control NOT Gates with controls $[x_i, y_i, a_0$ to $a_{m-j-1}]$
     - else
       - Apply multi control NOT Gates with controls $[x_i, y_i]$
     - end-if else
     - target → $[a_{m-j}]$
   - end-for
7) end-for
8) Measure Register $Q_a$

**Algorithm 1:** - Quantum Implementation of $a = \#_{1's}(\vec{x} \cap \vec{y})$

Algorithm 1 is implemented on the **IBM composer [7]**. IBM quantum composer is a virtual quantum simulator where we can execute quantum algorithms developed by IBM. Each qubit will have |0⟩ initially. Step 3 in Algorithm 1 will initialize three registers, $Q_x^{\otimes N} Q_y^{\otimes N}$ for two input vectors and $Q_a^{\otimes m}$ for storing the value **a** in binary format. As all qubits will have value |0⟩, we need to flip to |1⟩, when $x_i$ or $y_i = 1$, this happens in Step-4. In the Step 6, will the help of multicontrol NOT gates, value of register $Q_a$ will be incremented by value '1' if $x_i$ and $y_i = 1$. Then the required answer will be stored in register $Q_a$ and this is measured.

2) **Quantam Method for $\#_{1's}(\vec{x} \cup \vec{y})$:**

Let $\vec{x}$ and $\vec{y}$ are two binary vectors, $\vec{x}, \vec{y} \in \mathbb{R}^N$, where we need to find $\#_{1's}$ in Union between them. Prepare the state with three registers.

$$|\varphi\rangle = |x\rangle \otimes |y\rangle \otimes |b\rangle \qquad (6)$$

Where $|x\rangle = |x_{N-1} x_{N-2} x_{N-3}....x_0\rangle = |x_{N-1}\rangle|x_{N-2}\rangle|x_{N-3}\rangle...|x_0\rangle = |x_{N-1}\rangle \otimes |x_{N-2}\rangle \otimes |x_{N-3}\rangle...\otimes|x_0\rangle$.

$|y\rangle = |y_{N-1} y_{N-2} y_{N-3}.... y_0\rangle = |y_{N-1}\rangle|y_{N-2}\rangle|y_{N-3}\rangle...|y_0\rangle = |y_{N-1}\rangle \otimes |y_{N-2}\rangle \otimes |y_{N-3}\rangle...\otimes|y_0\rangle$.

A binary vector $\vec{b} \in \mathbb{R}^m$ can be used to store decimal values of $\#_{1's}$ in the Union, $\vec{b} = |b_m b_{m-1}...b_0\rangle$ and we get the integer value by

$$b = b_m 2^m + b_{m-1} 2^{m-1} .... + b_0 2^0; \quad (m = \lfloor \log N + 1 \rfloor) \qquad (7)$$

Apply algorithm 2 Union $(\vec{x}, \vec{y}, \vec{a})$, and measure the register $|b\rangle^{\otimes m}$

Union $(\vec{x}, \vec{y}, \vec{b})$:

1) **Input:** Vectors $\vec{x} = (x_{n-1}x_{n-2}...x_0)$, $\vec{y} = (y_{n-1}y_{n-2}....y_0)$
2) **Output:** $b = \#_{1's}(\vec{x} \cup \vec{y})$
3) Initialize the registers $Q_x^{\otimes N} Q_y^{\otimes N} Q_b^{\otimes m}$
4) Intersection $(\vec{x}, \vec{y}, \vec{a})$:
5) for i=0 to N-1:
   - apply CNOT gates with control $x_i$ and target $y_i$
6) end for
7) for i=0 to N-1:
   - for j=0 to m:
     - if (m-j-1) ≥ 0:
       - Apply multi control NOT Gates with controls $[y_i, a_0$ to $a_{m-j-1}]$
     - else:
       - Apply CNOT gate with control $y_i$
     - end-if else
     - target → $[a_{m-j}]$
   - end-for
8) end-for
9) Measure register $Q_b$

**Algorithm 2:** - Quantum Implementation of $b = \#_{1's}(\vec{x} \cup \vec{y})$

Algorithm 2 is implemented on **IBM composer.** Each qubit will have |0⟩ initially. Step 3 in Algorithm 1 will initialize three registers, $Q_x^{\otimes N} Q_y^{\otimes N}$ for two input vectors and $Q_b^{\otimes m}$ for storing the value **b** in binary format. Apply algorithm 1 intersection. Now we have $\#_{1's}(\vec{x} \cap \vec{y})$ in the register $Q_b$ after step 4. Then we add the cases when $x_i = 1$ and $y_i = 0$ (vice versa). In Step 5, CNOT gates essentially flips the $y_i$ qubit such that

```
if x_i = 1, y_i = 1:
        y_i = 0;
else-if x_i = 0, y_i = 1:
        y_i = 1;
else-if x_i = 1, y_i = 0:
        y_i = 1;
else    x_i = 0, y_i = 0:
        y_i = 0.
```

In the step 7, if $y_i = 1$; then the content of register $Q_b$ is incremented. The required answer will be stored in register $Q_b$ and this is measured.

From equation (1),

$$J(\vec{x}, \vec{y}) = \frac{a}{b} \qquad (8)$$

**Results**

We consider an example where $\vec{x} = 1010$ and $\vec{y} = 1101$ with N=4 and m= $\lfloor \log N + 1 \rfloor$ = 3. Algorithm 1 gives us $\#_{1's}$ which is the intersection of $\vec{x}, \vec{y}$ in $\vec{a}$ vector in binary format and Algorithm 2 gives us $\#_{1's}$ which is the union of $\vec{x}, \vec{y}$ in $\vec{b}$ vector in binary format.

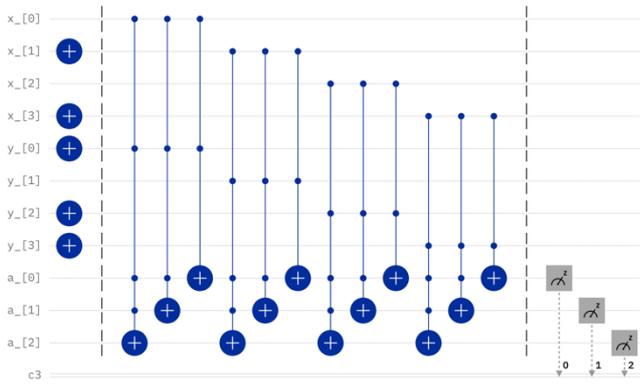

**Figure-1** Implementation of $\#_{1's}(\vec{x} \cap \vec{y})$ with $\vec{x}$ = 1010 and $\vec{y}$ = 1101 in IBM composer

Figure – I shows a quantum circuit, which is implemented on the IBM Quantum Composer to implement Algorithm 1. Three qubit registers are initialized, x_ [4] for $\vec{x}$, y_ [4] for $\vec{y}$ and a_ [3] register which we used to store the output vector i.e., $\vec{a}$. Apply the X gate if $x_i$ or $y_i = 1$. If we observe the first set of the multi control NOT gate controlling $x_0$, $y_0$, $a_0$, $a_1$ and targeting $a_2$, $a_1$, $a_0$, as $x_0 = 0$ there will be no change in the a_ register. In the next iteration $x_1$, $y_1$ will be controlled as $y_1 = 0$, and there will be no change in a_ register. Next iteration will control $x_2$, $y_2$ as $x_2 = 0$, and there will no change in a_ register. In the last iteration $x_3$, $y_3 = 1$ in the first iteration of the sub loop $a_1$, $a_0 = 0$, there will be no change in the $a_2$ qubit. As $a_0 = 0$ there will be no change in the $a_1$ qubit. In the last iteration of the subloop $a_0$ is controlled by $x_3$, $y_3$ which are equal to 1, and then $a_0$ flips to 1. Basically, when both $x_i$ and $y_i$ are 1 then register a will be incremented by "1" in the decimal value.

After Algorithm 1 has executed, Figure-3 show the result of register a_[3] i.e., $\vec{a}$ for the quantum circuit shown in figure-1, which is executed on the quantum simulator_statevector with 32 qubits on IBM composer for 1024 shots, the outcome $\vec{a}$ = 001, which is equivalent to a decimal value a = 1 by equation (5). The x-axis represents the frequency and y axis is results of the classical register we measured in Algorithm 1

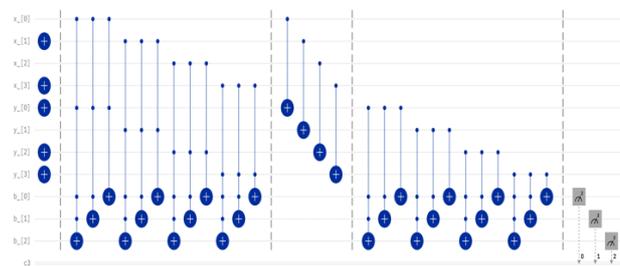

**Figure-2** Implementation of $\#_{1's}(\vec{x} \cup \vec{y})$ with $\vec{x}$ = 1010 and $\vec{y}$ = 1101 in IBM composer

In Figure – 2, there is a quantum circuit, which is implemented on IBM Quantum Composer, to implement Algorithm 2. There are three qubit registers x_ [4] for $\vec{x}$, y_ [4] for $\vec{y}$ and b_ [3] for $\vec{b}$. Then apply algorithm *Intersection* $(\vec{x}, \vec{y}, \vec{a})$. The result will store 001 in b_ register. Then apply CNOT gate with control $x_0$ and target $y_0$ as $x_0 = 0$, and there is no change in $y_0$. This step is repeated for every qubit in x and y registers. Now register y_ is with 1110, Then in first iteration of Step 7 in **algorithm 2**, and in first iteration in nested for loop $y_0$, $b_0 = 1$, $b_1 = 0$; no change in the $b_2$ qubit, second iteration in nested for loop $y_0$, $b_0 = 1$ then $b_1$ will flip to 1. Third iteration in nested for loop $y_0 = 1$, then the $b_0$ flips from 1 to 0. So, after the first iteration the b_ register value is 010 which is essentially equal to 2 in decimal value. In the same manner it takes all the iterations whenever it sees $y_i$ is 1 then b register will increment by 1.

After Algorithm 2 has executed, Figure-4 show the result of register b_[3] i.e., $\vec{b}$ for the quantum circuit figure-2 which is executed on quantum simulator_statevector with 32 qubits on IBM composer for 1024 shots, the outcome $\vec{b}$ = 100, which equivalent to decimal value b = 4 by equation (7). The x-axis represents the frequency and y axis is results of the classical register we measured in Algorithm 2

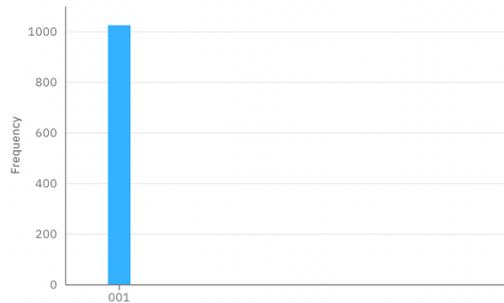

**Figure-3** Output for the quantum algorithm-1 in IBM simulator_statevector

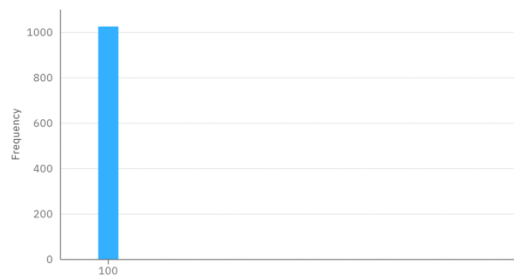

**Figure-4** Output for the quantum algorithm-2 in IBM simulator_statevector

## Conclusions

This paper proposes a quantum algorithm for the Jaccard Similarity. The algorithm has been implemented on the IBM composer which is a Quantum Computer simulator. Two sub algorithms have been proposed. Algorithm 1 give the result of intersection in a binary format $\vec{a}$ which is shown in Figure 3 with the implementation done on the Quantum Simulator, Algorithm 2 gives the result of union in binary format $\vec{b}$ which is shown in Figure 4. The proposed work measures only the numerator and denominator of the Jaccard Similarity. This can be extended in the terms of classification of data which can be represented as the binary format and applied to Quantum Machine algorithms such as KNN and Kmean's clustering. Superposition states can be added in order to reduce the qubit count and execute much faster. For each result we can map to a 1 hot vector which will be unique and can be applied in Quantum neural networks